# Atomistic insights on prebiotic phosphorylation of methanol from Schreibersite (Fe$_2$NiP) corrosion: *ab-initio* computational study

*Stefano Pantaleone,** [1] *Giulia De Gasperis,* [1] *Marta Corno,* [1] *Albert Rimola,* [2] *Nadia Balucani,* [3,4,5] *and Piero Ugliengo** [1]

[1] Dipartimento di Chimica and Nanostructured Interfaces and Surfaces (NIS) Centre, Università degli Studi di Torino, via P. Giuria 7, I-10125, Torino, Italy

[2] Departament de Química, Universitat Autònoma de Barcelona, 08193 Bellaterra, Catalonia, Spain

[3] Dipartimento di Chimica, Biologia e Biotecnologie, Università degli Studi di Perugia, Via Elce di Sotto 8, I-06123 Perugia, Italy

[4] Osservatorio Astrofisico di Arcetri, Largo E. Fermi 5, I-50125 Firenze, Italy

[5] Université Grenoble Alpes, CNRS, Institut de Planétologie et d'Astrophysique de Grenoble (IPAG), F-38000 Grenoble, France

**Abstract**

The prebiotic history of phosphorus is a matter of debate in the scientific community: its origin, how it landed on Earth, the selective speciation of the phosphate, and its inclusion into the organic matrix are the main unsolved issues. In this regard, Schreibersite (($Fe,Ni)_3P$), a mineral present in iron meteorites, can play a fundamental role as a carrier of reactive P which, as a result of the weathering processes, produces oxygenated phosphorus compounds, even the phosphate among others. In the present paper, we studied the interaction of methanol (alone and mixed with water) with the Schreibersite surfaces throughout periodic density functional theory calculations at PBE level. The results indicate that Schreibersite promotes the deprotonation of methanol and water both from thermodynamic and kinetic points of view, thus enabling the first step towards corrosion. We have simulated advanced stages of the corrosion process up to the formation of the phosphate and the phosphorylated form of methanol (methyl phosphate), showing that the formation of both products is thermodynamically favoured, as well as its solubilization, which allows other water molecules to proceed with further corrosion of Schreibersite.

# 1 – Introduction

The origin of life represents an issue very far to be solved in science, a debate born in the ancient Greece, and possibly even before, with the theories about spontaneous generation attributed to Aristotle,[1] which survived until the 19th century, when Stanley Miller and Joan Oró laid the foundations of modern prebiotic chemistry through their milestone experiments.[2,3] The formation of the simplest bricks of life (amino acids, nitrogenous bases, sugars, etc.) could occur either on the early Earth, as confirmed by the above-mentioned experiments, or in the interstellar medium throughout a variety of different reactions (radical-radical, photoinduced, surface catalysed) under so extreme conditions in terms of pressure and temperature to challenge our knowledge of chemical processes.[4]

Most of the studies on prebiotic chemistry focus their attention on the biogenic macro elements (SONCH) leaving aside phosphorus (P) which, despite representing a small fraction with respect to the other elements (1% w/w), is ubiquitous and essential in all living beings, through its presence in ATP, phospholipids as well as a part of the backbones of RNA and DNA where phosphate groups alternate sugar (ribose or deoxyribose) molecules. The most interesting questions in prebiotic chemistry concern the original reservoir of P, how the phosphate derived from that source, and which reactions lead to phosphate inclusion into the biological molecules.

The major reservoir of P on Earth is the apatite mineral ($Ca_5(PO_4)_3$[F, OH, Cl]). Nevertheless, apatite is very poorly soluble ($K_{ps} \sim 5\times10^{-58}$)[5] and reactive and, therefore, phosphate ($PO_4^{3-}$) groups are trapped in the inorganic matrix and not available to be incorporated into biomolecules. Indeed, phosphorylation of organic molecules, is a necessary step toward more complex bricks of life (e.g. nucleotides, phospholipids, the backbone of nucleic acids by its polymerization with ribose) [6]. Phosphorous-based materials may also help bypassing the "water paradox". This problem is a common thread of self-assembly that brings a fundamental concern: even though it is commonly accepted that water is a fundamental ingredient for the emergence of life, self-assembly that leads from monomeric units to oligo-/polymers and finally to macromolecules is based on condensation reactions in which water is one of the products and, therefore, an aqueous environment inhibits the reaction according to thermodynamics.[7] Indeed, condensed/poly phosphates ($P_4O_{10}$) materials may act as dehydrating agents during the phosphorylation processes.[8–11]

Another possibility, which is the one studied in the present paper, is that phosphorylation is promoted by reactive compounds where phosphorous is in a reduced state.[12] It is known that the most stable P oxidation state at the environmental conditions of the Earth crust is +5, *i.e.* the form of phosphate, mostly trapped into rocks, as already commented. Therefore, invoking the role of reduced P compounds implies to look outside Earth and much before its formation. According to most accepted theories, indeed, the intense meteor activity during the Archean era (4.0-3.8 billion years ago) could have been a major source of Schreibersite ($Fe_2NiP$),[13–16] which is one of the minor phases of iron meteorites.[17–20] Interestingly, it has been recently demonstrated that in situ synthesis of Schreibersite is possible by lightning strikes hitting clay rich soils, thus reinforcing the possible role of this mineral even without any constrain on the relatively short time window of the late meteoritic bombardment. According to this study, lightning strikes on early Earth could potentially have formed 10–1000 kg/y of phosphide and 100–10000 kg/y of phosphite and hypophosphite.[21]

The first redox experiment carried out on Schreibersite by Pasek and coworkers[22] demonstrated for the first time the possible prebiotic role of this mineral because its corrosion by water leads to the formation of several different P oxygenated compounds, including phosphates. In subsequent experiments the phosphorylation of two nucleosides was achieved by simply mixing and heating an aqueous solutions containing the organic compounds in the presence of synthetic analogues of the meteoritic schreibersite.[23] Furthermore, by probing the ability of Schreibersite to phosphorylate other organics, it was found that it catalyzes the aldol reaction of small carbohydrates forming larger sugar molecules,[23,24] thus implying a multiple fundamental role of Schreibersite in the prebiotic context.[24]

The adsorption of water and simple organic molecules such as methanol and formic acid on synthetic Schreibersite was recently studied by means of RAIRS (Reflection Absorption InfraRed Spectroscopy) at low temperatures (120 K). The study shows a possible spontaneous methanol dissociation, at variance with what happens to water and formic acid, thus further enhancing the interest to study the former molecule over the others,[25] because methanol can be thought as a crude approximation of a sugar like ribose, at least for what concern the reaction pathway that leads to the corresponding phosphorylated product.

Very few computational works have been found in literature, and the majority of them is devoted to study the phase stability of Schreibersite in planetary cores under high pressure and temperature conditions.[26–28] Recent papers, mainly coming from our research groups, studied the properties of bulk and surfaces of Schreibersite in interaction with water,[29,30] as well as its reactivity towards water corrosion.[31,32] In particular, in the last paper published by some of us, we clearly show that the formation of the phosphate from physisorbed water at the Schreibersite surfaces is a strongly favoured process from a thermodynamic point of view and, at the same time, the barrier accounting for the deprotonation of water is very low, resulting in instantaneous reactions at room temperature.

In this work, the adsorption and reactivity of methanol and mixtures of methanol and water on the (110) and (001) Schreibersite surfaces were studied using periodic theoretical simulations at Density Functional Theory (DFT) level with the aim to understand interface interactions and derived physico-chemical outcomes at the atomistic level.

## 2 - Computational Details

The adsorption of methanol and methanol/water on Schreibersite was studied by means of periodic DFT calculations carried out with the Vienna Ab-initio Simulation Package (VASP) code,[33–36] which uses projector-augmented wave (PAW) pseudopotentials[37] to describe the ionic cores and a plane wave basis set for the valence electrons. Geometry optimizations and frequency calculations were performed with the gradient corrected PBE functional,[38] with a posteriori Grimme D2 correction,[39] modified for solids (D*).[40] Moreover, the C6 atomic coefficients related to polarizabilities on Fe and Ni metal atoms were set to 0 (*i.e.* no dispersion interaction contribution from metal atoms). On O, C and H atoms the original D* parameters were used. This method of choice is referred to as PBE-D*0 along the work. This setup was chosen according to the best results obtained in our previous work on the bulk and bare surfaces of Schreibersite.[29] The cutoff energy of plane waves (which control the accuracy of the calculations) was set to 500 eV. The self-consistent field (SCF) iterative procedure was converged to a tolerance in total energy of $\Delta E = 10^{-5}$ eV for geometry optimizations, while for frequency calculations the tolerance was decreased to $\Delta E = 10^{-6}$ eV. As Schreibersite is a magnetic material, all calculations were carried out with the spin-polarization active.

The study has considered the (110) and (001) Schreibersite surfaces, as they are the most and least stable ones, respectively. All details about the surface modelling are available in Ref. [29]. The Monkhorst-Pack sampling of the Brillouin zone was used for the *k*-points mesh. Shrinking factors for the (110) surface have been set on (8 8 1) on the unit cell (a = 4.374 Å, b = 6.719 Å, γ = 108.998°, surface thickness equal to 17.0 Å), and to (4 4 1) for 2x1 and 2x2 supercell models (for a total number of 64 and 16 *k*-points, respectively), while for the (001) surface (a = 8.984 Å, b = 8.984 Å, γ = 90.000°, surface thickness equal to 13.2 Å) they have been set to (4 4 1) (16 *k*- points). The shrinking factor related to the non-periodic direction was always set to "1" (*i.e.* no sampling of the reciprocal space). As VASP relies on plane waves basis set, surfaces are replicated even along the non-periodic direction, the vacuum space among fictitious replicas being set to at least 20 Å to minimize the interactions among replica images. Therefore, the final *c* cell axis was set to 40 Å. Geometry relaxations were carried out by moving all atoms in the unit cell (all the atomic layers were set free to relax) while keeping the cell parameters fixed at the geometry optimized for the bare surface to enforce the rigidity due to the underneath bulk. For transition state search the DIMER method was adopted,[41–44] while for updating the Hessian during geometry optimizations (both for minima and transition states) the Conjugate Gradient algorithm has been employed. The tolerance on gradients during the optimization procedure was set to 0.01 eV/Å for each atom in each direction.

Adsorption energies (AE) of methanol at the Schreibersite surfaces were calculated as:

$$AE = \frac{E_{CPLX} - (nE_{CH_3OH} + E_{surf})}{n} \qquad Eq. 1$$

where $E_{CPLX}$ is the potential energy of the complex (methanol adsorbed on the surface), $E_{CH_3OH}$ and $E_{surf}$ are the potential energies of the isolated methanol molecule and the (110) or (001) bare surfaces, each one calculated at its relaxed geometry, while *n* is the number of adsorbed methanol molecules per surface unit cell. When considering methanol dissociation, the reaction energy is computed with respect to the most stable molecular adsorption on the same surface and labelled $\Delta E$ (deprotonation/dissociation energy). Zero Point Energy (ZPE) and thermal contributions to the vibrational energy were included to calculate the reaction enthalpy ($\Delta H$), to which the vibrational entropy was added thus obtaining the reaction Gibbs free ($\Delta G$). The rotational and translational contributions were only considered for the free $CH_3OH$ and $H_2O$ molecules.

The kinetic rate constant of methanol dissociation, was computed using the Eyring's transition state theory equation assuming a unimolecular process:[45,46]

$$k = \frac{k_B T}{h} e^{\frac{\Delta G^{\ddagger}}{RT}} \qquad Eq.2$$

where $k_B$ is the Boltzmann constant, $T$ is the absolute temperature, $R$ is the ideal gas constant, h is the Planck constant and $\Delta G^{\ddagger}$ is the difference of Gibbs energy between the transition state and the previous corresponding minimum. The half-life time $t_{1/2}$ has been estimated assuming first-order kinetics, as:

$$t_{1/2} = \frac{ln2}{k} \qquad Eq.3$$

As regards the deprotonation reactions, vibrational frequencies were computed at Γ point on a reduced Hessian (only the atoms of methanol were set free to move), by numerical differentiation of the analytical first derivatives, using the central difference formula (*i.e.* two displacements of 0.02 Å for each atom in each (x,y,z) direction), in order to confirm that the optimized structure is a minimum (all real frequencies) or a first order saddle point (all real frequencies but one imaginary frequency). The Phonopy[47] code was used for both generating atomic displacements and processing VASP outputs. Thermodynamic corrections to the energy were calculated trough the Quasi-Harmonic approximation as proposed by Grimme,[48] in which frequencies lower than 100 cm$^{-1}$ are replaced by free rotor modes when calculating the entropy. This improves the calculation of the vibrational entropy, which would be underestimated when considering very low frequency values. To avoid discontinuities close to the cut-off value, a damping function was used to interpolate the values of entropy computed with the two approaches. Visualization and manipulation of the structures and figures rendering was done with the MOLDRAW,[49] VMD,[50] and POVRAY[51] programs.

## 3 – Results and Discussion

*3.1 – Methanol physisorption and chemisorption at (110) Schreibersite surface*

The physisorption of methanol (i.e. the adsorption of molecular $CH_3OH$) was modelled starting from the optimized structures of adsorbed water from our previous paper[30] and replacing the farthest H atoms to the surface with the $CH_3$ group minimizing the steric hindrance.

Figure 1 shows methanol physisorption on the unit cell of the Schreibersite (110) slab, and the corresponding adsorption energies. As for water, the most stable adsorption mode of methanol is on Ni (AE = -42.5 kJ/mol, see Eq.1) with a slightly larger adsorption energy than water, possibly due to the higher dispersion interactions exerted by the -$CH_3$ group with respect to a single hydrogen atom. The small size of the unit cell does not allow any other adsorption site than Ni and Fe. As for water, no direct interaction with P is possible for the molecular methanol, because of the repulsive force between the partially negative charges on both the O of methanol and the P of Schreibersite. The effect of the surface coverage is like that of water (see Table S1 and Figure S1), *i.e.* from a structural point of view the O—metal bond shortens moving from high to low methanol loading, with a corresponding decrease in the adsorption energy, with a stronger effect on Ni (from -42.5 to -55.3 kJ/mol) than on Fe (from -28.1 to -35.5 kJ/mol).

At variance with the physisorption, methanol chemisorption presents several possibilities, depending on the starting molecular site (Ni or Fe), and on the target of the displaced proton (see Figure 2). Here we studied the $CH_3OH$ chemisorption by manually breaking the OH bond towards surface protonation and letting the structure to fully relax. To ensure an exhaustive sampling of methanol dissociation, a 2x1 supercell was adopted (a = 8.748 Å, b = 6.719 Å, γ = 108.99°), and the H atom of the OH group was manually displaced from methanol taking into account all nearby surface atoms (Fe, Ni, P). In this case, even P becomes an active site, either for the proton or for the methoxy group, whose oxygen atom misses a bond and is consequently prone to oxidize the P atom. However, this structure (Figures 2e) has an endoergic deprotonation energy (DE = 36.8 kJ/mol), calculated with respect to the most stable adsorption mode (110-meth-Ni_1x1). On the contrary, the most stable deprotonation occurs on the metallic sites, the most stable ones being the cases in which both the oxygen of the $CH_3O^-$ group and $H^+$ are shared among more than one metal center (Figure 2a, DE = -18.5 kJ/mol). Indeed, in our previous paper concerning the interaction of water with Schreibersite, we provided a ranking of stabilization for the redox reaction of the outermost atoms of the surface when they get in contact with water. According to our previous results, the first atomic species to undergo redox reactions are the metals (Fe and Ni), which are almost indistinguishable either concerning the H or OH attack, then the hydroxylation of P, and finally the hydriding of phosphorus. Here, the trend is similar for methanol, where $CH_3O$ plays the same role as OH for water.

Figure 3 reports the mechanism of methanol dissociation on the (110) 2x1 surface, *i.e.* the first step of phosphorylation, including the energy barrier ($\Delta E^\ddagger$ = 92.7 kJ/mol, $\Delta G^\ddagger_{298}$ = 73.1 kJ/mol), while Table S2 and Figure S2 show the details about reaction thermodynamics and kinetics at different temperatures (see Eq. 2 and 3). At 298 K, the reaction is very fast ($t_{1/2}$=3×10$^{-4}$ hours, less than 1 second) and exergonic ($\Delta H$ = -18.5 kJ/mol, $\Delta G_{298}$ = -22.1 kJ/mol), indicating that even the most stable surface promotes the deprotonation and, possibly, even phosphorylation.

When dealing with water adsorption, we found out that all deprotonation reactions were endoergic (the most stable one presenting a deprotonation energy of 29.1 kJ/mol). To assess why the two cases are different, we started from the most stable deprotonated structures of methanol and replaced the $CH_3$ group by a H atom, thus going back to the water case. A new stable structure was found, in which the reaction is now exoergic ($\Delta E$ = -25.1 kJ/mol, see Figure S3). We can therefore conclude that both Schreibersite surfaces undergo corrosion in a similar way for the two molecules (water and methanol), as expected by their similar acidity (pKa = 15.7 and 15.5, for water and methanol, respectively).

*3.2 – Methanol physisorption and chemisorption at (001) Schreibersite surface*

Figure 4 shows the molecular adsorption of methanol on the Schreibersite (001) 1x1 reactive surface (a = 8.984 Å, b = 8.984 Å, γ = 90.0°). Due to the higher surface energy of the (001) surface, we expect lower adsorption energies with respect to the (110) surface. Indeed, this is the case, where Fe is the preferred site (001-meth-Fe, AE = -71.3 kJ/mol). The (001) surface presents a much more corrugated shape than the (110), and the most stable adsorption occurs at the most protruding (and less coordinated) sites, *i.e.* Fe (Figure 4a). The innermost surface atoms exhibit, instead, comparable adsorption energies with the (110) surface, even in the case of bidentate adsorption mode (Figure 4b, 001-meth-FeFe, AE = -44.6 kJ/mol).

The dissociation of methanol is reported in Figure 5, using the same unit cell as for the molecular adsorption, as in this case the distance among adsorbate replicas is far enough to ensure negligible lateral interactions. The dissociation energies on the (001) surface are much larger than those on the (110) (001-meth-dep1, $\Delta E$ = -92.3 kJ/mol), even if the priority of the atoms undergoing corrosion is similar to those of the most stable surface. In this case, we explored another methanol dissociation channel, *i.e.* the simultaneous breaking of the O—H and C—O bonds (Figure 5a), which represents the most stable structure; this is not unexpected, as there are more atoms saturating and, accordingly, stabilizing the undercoordinated species of the surface. This dissociation can be of particular interest for prebiotic processes with respect to phosphorylation (the focus of this work) because of the presence of a direct chemical bond between carbon and Fe/Ni. This case envisages organometallic chemistry, as C has a carbanion character, which is the ideal situation for a nucleophilic attack to a partially positive C (a much more common electronic situation in most organic moieties), thus giving the possibility to elongate carbon chains, another fundamental reaction in biochemistry. Indeed, Bader analysis confirmed that the charge on the C atom is -0.55 $e$. However, this is not the scope of the present work and, therefore, will not be further analyzed.

Finally, the reaction mechanism of methanol dissociation on the (001) surface was also studied (see Figure 6). As for the binding and deprotonation energies, which are more favorable on the (001) surface, also the activation barrier is smaller ($\Delta E^{\ddagger} = 51.8$ kJ/mol, $\Delta G^{\ddagger}_{298} = 33.1$ kJ/mol) compared to the one on the (110) surface ($\Delta E^{\ddagger} = 92.7$ kJ/mol, $\Delta G^{\ddagger}_{298} = 73.1$ kJ/mol). In Table S3 and Figure S4 the rate constants and half-life times are reported, showing an instantaneous timescale of the reaction at room temperature as for the (110), even if on the (001) surface the reaction is orders of magnitude faster ($2 \times 10^{-11}$ hours, which corresponds to a timescale of microseconds).

*3.3 – Towards phosphorylation*

The focus of the present paper is to study the phosphorylation of organic molecules promoted by Schreibersite, because of its capability of complexing simple sugars, the most interesting one being ribose, whose phosphorylation leads to ribose phosphate. This species is of fundamental importance in biochemistry: as an example, it represents two out of three fragments of the AMP (adenosine monophosphate), while its polymer constitutes the backbone of the DNA double helix. To this end, we used the methanol molecule as a prototype of ribose, which was linked to the phosphate, as depicted in Figure 7c. As four oxygen atoms are necessary to obtain the phosphate ($PO_4^{3-}$), the reactant was modelled with one methanol and three water molecules physisorbed on the surface (Figure 7a), thus ensuring the mass balance with respect to the final product. Two different products were simulated: the first one with the phosphate species, as already presented in our previous work, and the $CH_3$ fragment on the surface (Figure 7b), and a second one with the phosphorylated methanol molecule. The results show that the most favorable product is the phosphate alone (*i.e.* without $CH_3$) ($\Delta E = -182.3$ kJ/mol), even if the phosphorylation is very close in energy ($\Delta E = -169.1$ kJ/mol) and therefore it likely represents a competitive channel.

As a further step, we also considered the possibility of solubilization of the phosphorylated product by simulating the adsorption of methoxyphosphonic acid ($CH_3OP(O)(OH)_2$) on a "micro-hydrated" surface (see Figure 8d, 001-prod-phospho-solution). To ensure the mass balance with respect to the reactant (001-reag, Figure 8b) and the phosphate product (001-prod-phosphorylation, Figure 8c) the reference was considered taking only non-interacting species, *i.e.* the Schreibersite surface, the water and methanol molecules as isolated, in order to include as many water molecules as needed to obtain meaningful energetic differences among structures that present a different number of atoms. The reaction of $3H_2O + CH_3OH + $ (001) Schreibersite leading to the methyl phosphate is strongly exoergic with respect to the isolated reactants ($\Delta E = -427.2$ kJ/mol), but the solubilization of the methyl phosphate is even more favorable ($\Delta E = -434.5$ kJ/mol), indicating that water molecules can bring the phosphorylated product into the solution. Due to the large size of these systems, in particular the last one (001-prod-phospho-solution), we carried out the simulation of the same product but on a larger 2x2 cell, in order to evaluate the effect of lateral interactions that in these models are expected to play an important role. The reactants are in this case $3H_2O + CH_3OH + $ 2x2 (001) Schreibersite and $6H_2O + CH_3OH + $ 2x2 (001) Schreibersite, for the methyl phosphate chemisorbed on the surface and the methyl phosphate physisorbed on a microsolvated Schreibersite surface, respectively. Both reactions are even more favorable by adopting the 2x2 unit cell. For 001-prod-phosphorylation_2x2 the reaction energy is $\Delta E$ = -489.9 kJ/mol, while the methyl phosphate solubilization (i.e. 001-prod-phospho_solution_2x2) becomes even more favorable ($\Delta E$ = -561.3 kJ/mol). The latter case was more sensitive to the cell size due to the higher number of water molecules with respect to the former. In summary, the calculations performed on the unit cell (i.e. with a higher coverage of water molecules) represent an upper limit of the energetics of the process, which becomes more favorable when enlarging the unit cell of the surface (i.e. at lower water coverage regimes). This confirms Schreibersite as very reactive component towards corrosion, so the possible mechanism is that phosphorus oxygenated compounds, as well as metal oxides and hydroxides, are continuously produced and released into the water solution up to the full corrosion of the material.

## 4 - Conclusions

In this work, the interaction of methanol with the (110) and (001) Schreibersite surfaces, physisorption and chemisorption, has been assessed by means of periodic density functional theory simulations at PBE-D*0/plane waves level with the VASP code. The main achievements are that either the (110) and (001) surfaces, *i.e.* the most and least stable ones, chemisorb methanol with the following priority: i) redox reaction on metals (Fe/Ni—H and Fe/Ni—OCH$_3$), ii) oxidation of P (P—OCH$_3$), iii) reduction of P (P—H). The first two reactions are exoergonic ($\Delta G_{298}^{(110)} = -22.1$ kJ/mol and $\Delta G_{298}^{(001)} = -64.8$ kJ/mol), with activation barriers for the methanol deprotonation rather small ($\Delta G_{298}^{\ddagger(110)} = 73.1$ kJ/mol and $\Delta G_{298}^{\ddagger(001)} = 33.1$ kJ/mol), corresponding to very fast rate constants (orders of seconds and microseconds, respectively), indicating that Schreibersite easily undergoes corrosion. As regards phosphorylation, the production of phosphate and methyl phosphate are both largely exoergic reactions, $\Delta E = -182.3$ kJ/mol and $\Delta E = -169.1$ kJ/mol, respectively. Finally, we evaluated the energetics of methyl phosphate solubilization, showing that it prefers to interact with a micro hydrated Schreibersite than a direct contact with the surface. These results confirm the high potential of Schreibersite as a key material to disentangle the long-standing issue about the availability of prebiotic phosphorus and its inclusion into organic matter, thus forming some fundamental bricks of life.

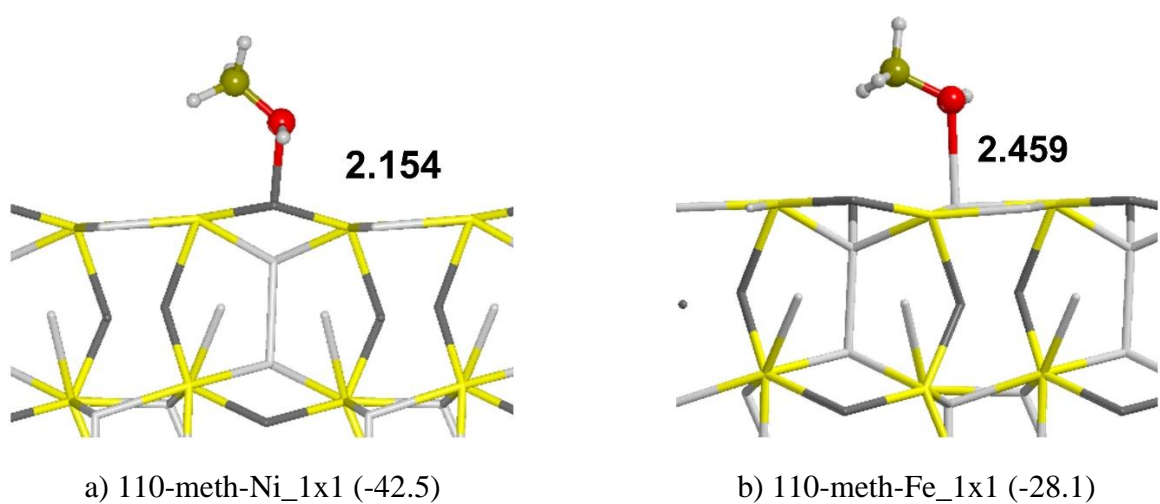

a) 110-meth-Ni_1x1 (-42.5)  b) 110-meth-Fe_1x1 (-28.1)

Figure 1. PBE-D*0 optimized structures of physisorbed methanol molecule on the (110) Schreibersite surface. Adsorption energies (values in parenthesis) are in kJ/mol. Distances are in Å. Atom color legend: H is in white, C in ochre, O in red, P in yellow, Fe in light grey, and Ni in dark grey.

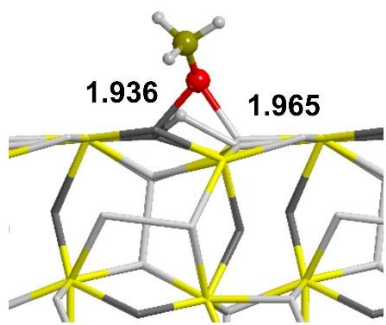 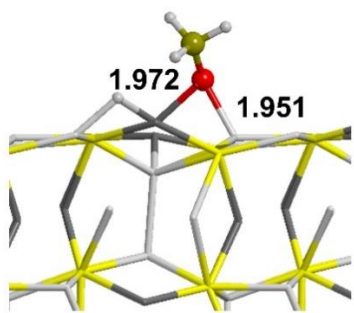 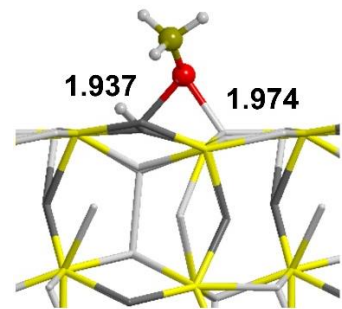

a) 110-meth-dep1 (-68.8) [-18.5]     b) 110-meth-dep2 (-39.0) [11.3]     c) 110-meth-dep4 (-55.4) [-5.1]

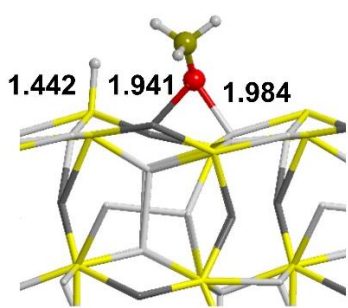 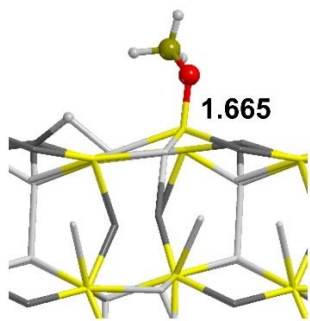

d) 110-meth-dep5 (3.8) [54.2]     e) 110-meth-dep7 (-13.4) [36.8]

Figure 2: PBE-D*0 optimized structures of chemisorbed methanol molecule on the (110) Schreibersite surface. Adsorption and deprotonation energies (values in round and square parenthesis) are in kJ/mol, taking as the $0^{th}$-energy reference the 110-meth-Ni_2x1 system, (*i.e.* the most stable methanol molecular physisorption). Distances are in Å. Atom color legend: H is in white, C in ochre, O in red, P in yellow, Fe in light grey, and Ni in dark grey.

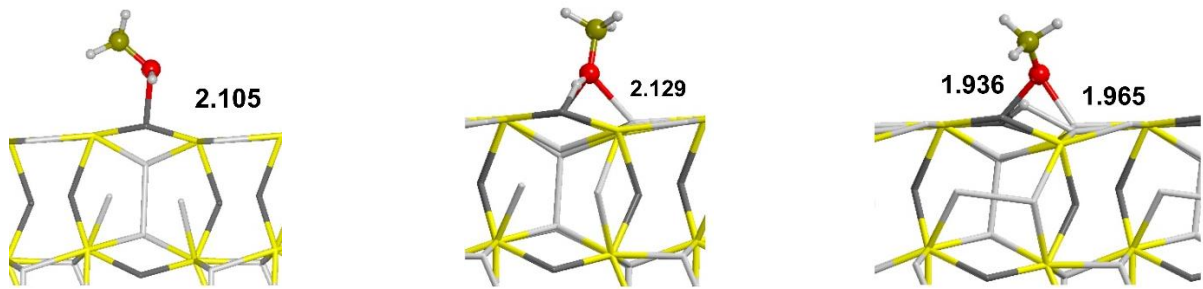

a) 110-meth-Ni_2x1 [0.0] {0.0}     b) 110-meth-TS [92.7] {73.1}     c) 110-meth-dep1 [-18.5] {-22.1}

Figure 3: PBE-D*0 optimized structure of methanol dissociation on the (110) Schreibersite surface. Reaction energies are in kJ/mol, taking the reactant as the 0$^{th}$-energy reference: in square parenthesis the reaction energies, in curly parenthesis the reactions free energies at 298 K. Distances are in Å. Atom color legend: H in white, C in ochre, O in red, P in yellow, Fe in light grey and Ni in dark grey.

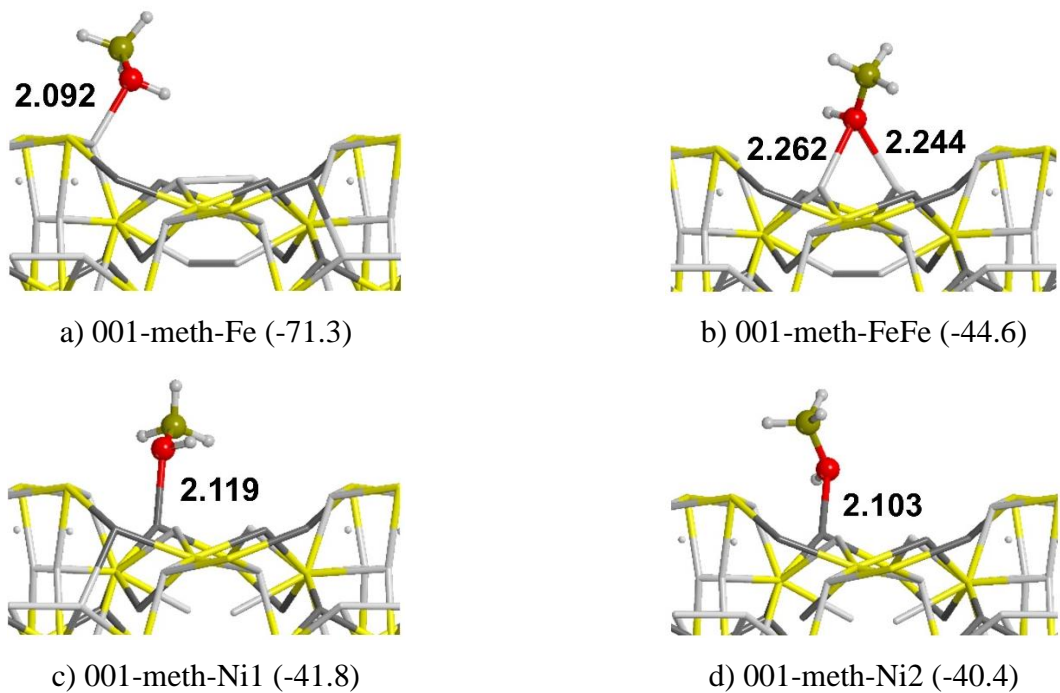

Figure 4: PBE-D*0 optimized structures of physisorbed methanol molecule on the (001) Schreibersite surface. Adsorption energies (values in parenthesis) are in kJ/mol. Distances are in Å. Atom color legend: H is in white, C in ochre O in red, P in yellow, Fe in light grey, and Ni in dark grey.

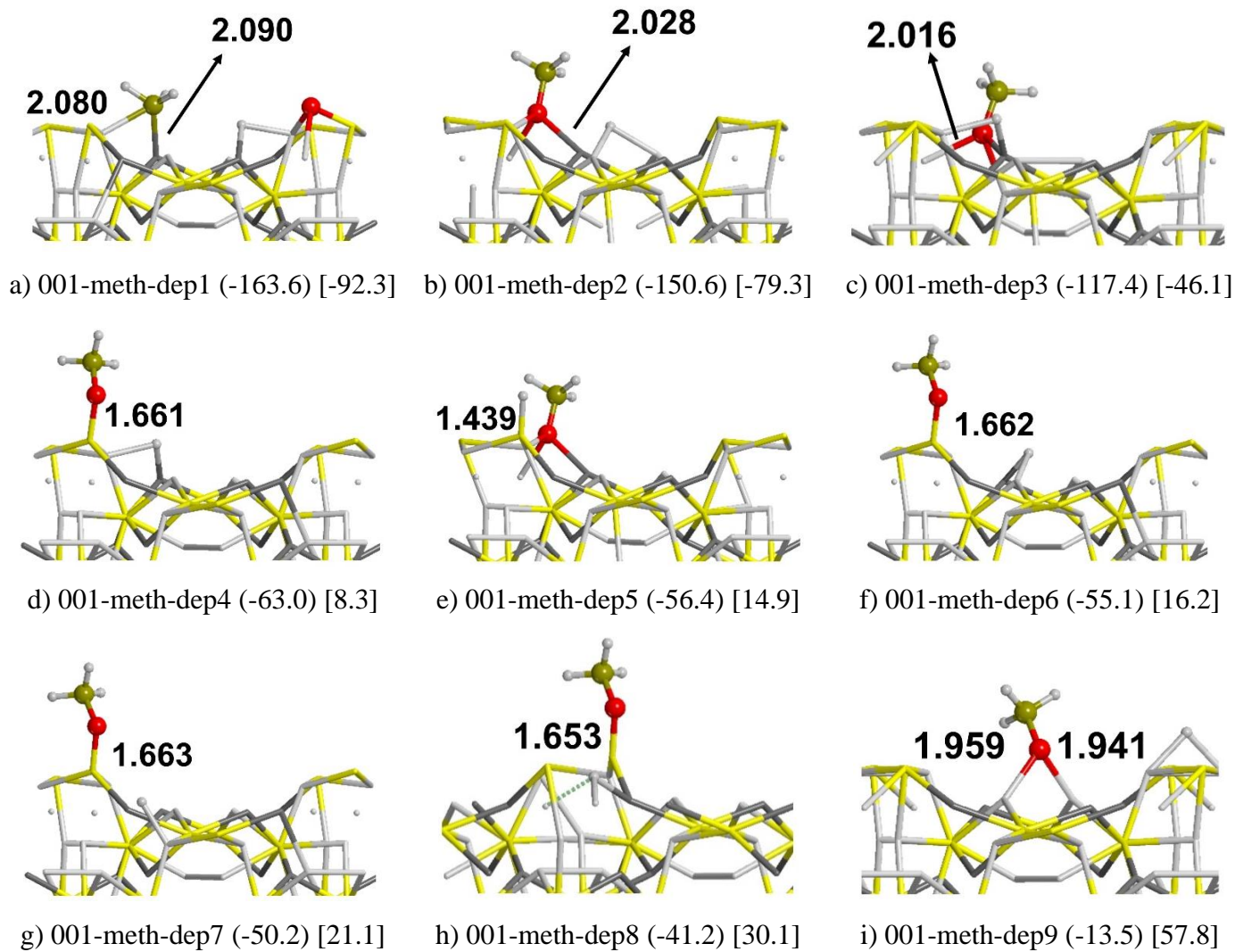

a) 001-meth-dep1 (-163.6) [-92.3]   b) 001-meth-dep2 (-150.6) [-79.3]   c) 001-meth-dep3 (-117.4) [-46.1]

d) 001-meth-dep4 (-63.0) [8.3]   e) 001-meth-dep5 (-56.4) [14.9]   f) 001-meth-dep6 (-55.1) [16.2]

g) 001-meth-dep7 (-50.2) [21.1]   h) 001-meth-dep8 (-41.2) [30.1]   i) 001-meth-dep9 (-13.5) [57.8]

Figure 5: PBE-D*0 optimized structures of chemisorbed methanol molecule on the (001) Schreibersite surface. Adsorption and deprotonation energies (values in round and square parenthesis) are in kJ/mol, taking as the $0^{th}$-energy reference the 001-meth-Fe system, (*i.e.* the most stable methanol molecular physisorption). Distances are in Å. Atoms color legend: H is in white, C in ochre, O in red, P in yellow, Fe in light grey, and Ni in dark grey.

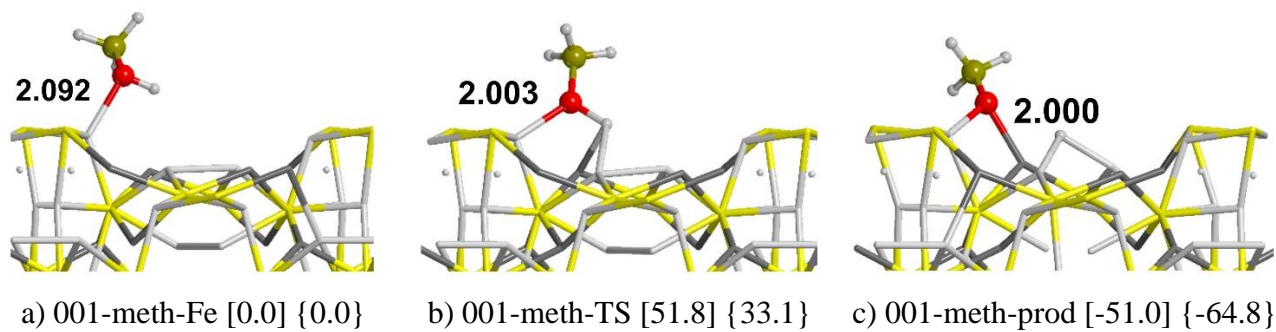

a) 001-meth-Fe [0.0] {0.0}   b) 001-meth-TS [51.8] {33.1}   c) 001-meth-prod [-51.0] {-64.8}

Figure 6: PBE-D*0 optimized structures of methanol dissociation on the (001) Schreibersite surface. Reaction energies are in kJ/mol, taking the reactant as the $0^{th}$-energy reference: in square parenthesis the reaction energies, in curly parenthesis the reactions free energies at 298 K. Distances are in Å. Atom color legend: H in white, C in ochre, O in red, P in yellow, Fe in light grey and Ni in dark grey.

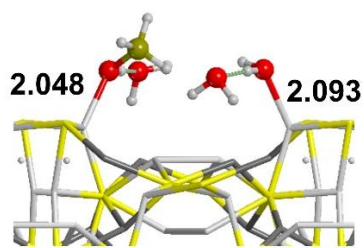 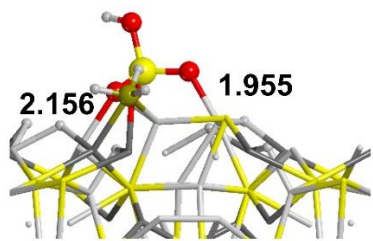 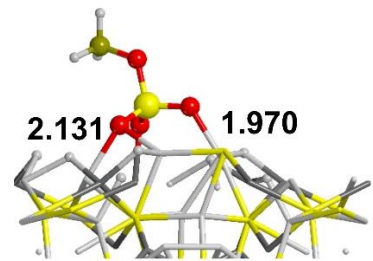

a) 001-reag [0.0]   b) 001-prod-phosphate [-182.3]   c) 001-prod-phosphorylation [-169.1]

Figure 7: PBE-D*0 optimized structures of reactant and products of one of the possible phosphorylation reaction channels. Reaction energies (values in brackets) are in kJ/mol, taking the reactant as the $0^{th}$-energy reference. Atom color legend: H in white, C in ochre, O in red, P in yellow, Fe in light grey and Ni in dark grey.

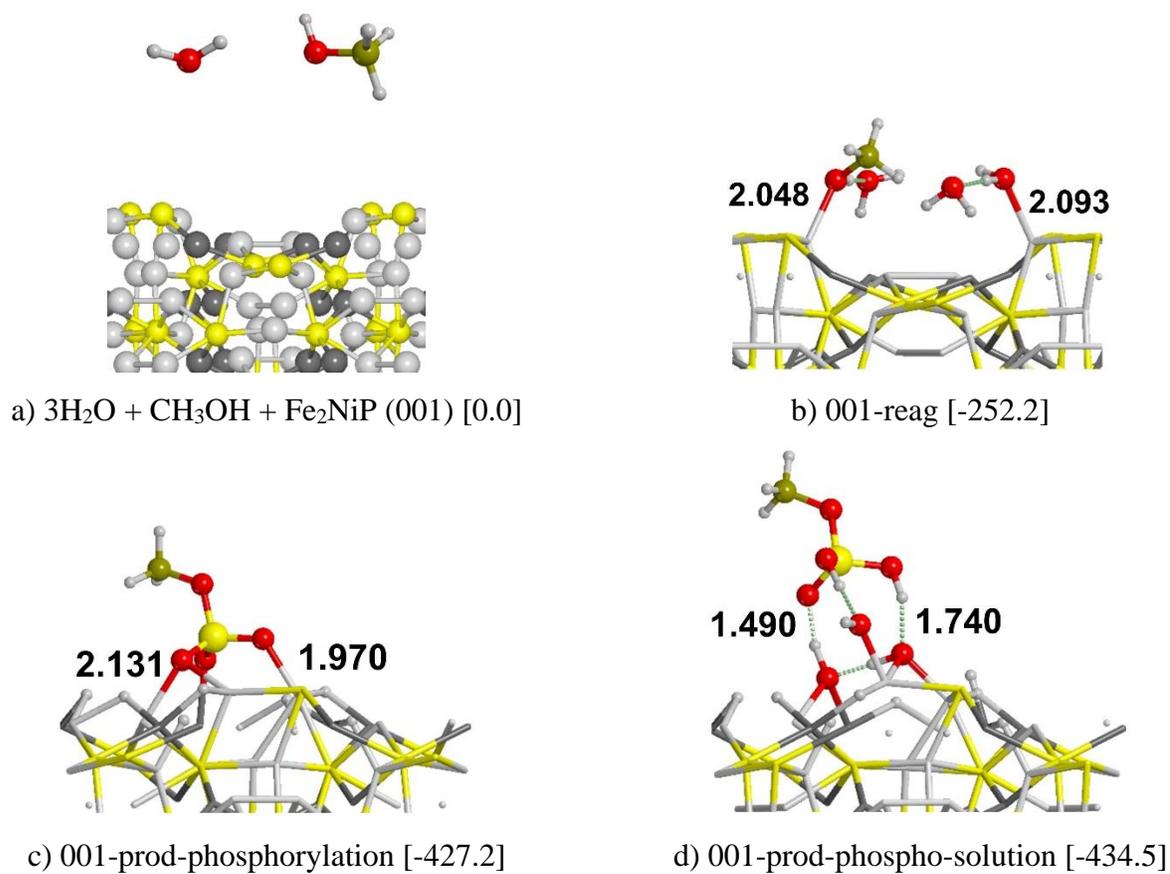

a) 3H₂O + CH₃OH + Fe₂NiP (001) [0.0]  
b) 001-reag [-252.2]  
c) 001-prod-phosphorylation [-427.2]  
d) 001-prod-phospho-solution [-434.5]

Figure 8: PBE-D*0 optimized structures of reactant and products of methyl phosphate solubilization. Reaction energies (values in brackets) are in kJ/mol, taking the reactant as the $0^{th}$-energy reference with all the isolated species, *i.e.* the surface plus water and methanol in the gas phase. Distances are in Å. Atom color legend: H in white, C in ochre, O in red, P in yellow, Fe in light grey and Ni in dark grey.

ASSOCIATED CONTENT

**Supporting Information**. The following files are available free of charge. PDF file content: A figure with more deprotonated structures and corresponding relative energy with respect to the most stable one, the coordinates of the optimized structures in POSCAR format and a .tar file with the optimized structures in POSCAR format as well.

AUTHOR INFORMATION

**Corresponding Author**

*To whom correspondence should be addressed:
stefano.pantaleone@unito.it, piero.ugliengo@unito.it

**Author Contributions**

The manuscript was written through contributions of all authors. All authors have given approval to the final version of the manuscript. The results reported in this paper have been partly obtained during the M.S. thesis of Giulia De Gasperis (University of Torino, 2022-23).

**Funding Sources**

This work has been partially supported by the Spoke 7 "Materials and Molecular Sciences" of ICSC – Centro Nazionale di Ricerca in High-Performance Computing, Big Data and Quantum Computing, funded by European Union – NextGenerationEU, from the Italian MUR (PRIN 2020, "Astrochemistry beyond the second period elements", Prot. 2020AFB3FX) is gratefully acknowledged and by the Italian Space Agency (Bando ASI Prot. n. DC-DSR-UVS-2022-231, Grant no. 2023-10-U.0 "Modeling Chemical Complexity: all'Origine di questa e di altre Vite per una visione aggiornata delle missioni spaziali (MIGLIORA)" Codice Unico di Progetto (CUP)

F83C23000800005). This project received funding from the European Research Council (ERC) under the European Union's Horizon 2020 Research and Innovation Program (Grant Agreement 865657) for Project "Quantum Chemistry on Interstellar Grains" (QUANTUMGRAIN). MICIN is also acknowledged for financing the projects PID2021-126427NBI00 and CNS2023-144902.

**Acknowledgments**

We acknowledge the EuroHPC Joint Undertaking for awarding this project access to the EuroHPC supercomputer LUMI, hosted by CSC (Finland) and the LUMI consortium through the EuroHPC Regular Access call n° EHPC-REG-2023R03-117, and the CINECA award under the ISCRA initiative, for the availability of high-performance computing resources and support.

The authors acknowledge support from the Project CH4.0 under the MUR program "Dipartimenti di Eccellenza 2023–2027" (CUP: D13C22003520001).